\documentclass[conference]{IEEEtran}
\IEEEoverridecommandlockouts

\usepackage{comment}
\usepackage{cite}
\usepackage{amsmath,amssymb,amsfonts}
\usepackage[english]{babel}
\usepackage[utf8]{inputenc}
\usepackage{graphicx}
\usepackage{algorithm}
\usepackage{array}
\usepackage{tabularx}
\usepackage[noend]{algpseudocode}
\usepackage{booktabs}
\usepackage{tabulary}
\usepackage{textcomp}
\usepackage{xcolor}
\usepackage{url}
\usepackage{subcaption}
\def\BibTeX{{\rm B\kern-.05em{\sc i\kern-.025em b}\kern-.08em
    T\kern-.1667em\lower.7ex\hbox{E}\kern-.125emX}}

\usepackage[colorlinks]{hyperref}
\usepackage{balance}

\usepackage{ifthen}
\usepackage[normalem]{ulem} 
\usepackage{xcolor}

\newboolean{showedits}
\setboolean{showedits}{true} 
\ifthenelse{\boolean{showedits}}
{
	\newcommand{\del}[1]{\textcolor{red}{\sout{#1}}} 
}{
	\newcommand{\del}[1]{} 
	
}

\newboolean{showcomments}
\setboolean{showcomments}{true}
\newcommand{\id}[1]{$-$Id: scgPaper.tex 32478 2010-04-29 09:11:32Z oscar $-$}

\ifthenelse{\boolean{showcomments}}
{\newcommand{\nbc}[3]{
 {\colorbox{#3}{\bfseries\sffamily\scriptsize\textcolor{white}{#1}}}
 {\textcolor{#3}{\sf\small$\blacktriangleright$\textit{#2}$\blacktriangleleft$}}}
 }
{\newcommand{\nbc}[3]{}
 \renewcommand{\del}[1]{} 
 }
 \definecolor{darkyellow}{RGB}{255, 222, 9}

\definecolor{ibcolor}{rgb}{0.4,0.6,0.2}
\definecolor{ascolor}{rgb}{0,0.5,0.9}

\definecolor{bocolor}{rgb}{0.6,0.9,0.2}
\definecolor{jrcolor}{rgb}{0.5,0,0.5}
\definecolor{nrcolor}{rgb}{0.4,0.1,0.3}
\definecolor{hkcolor}{rgb}{1.0,0.5,0.3}
\definecolor{tdcolor}{rgb}{1.0,0,0}

\newcommand\iv[1]{\nbc{IB}{#1}{ibcolor}}

\definecolor{freakishgreen}{HTML}{0A982B}

\hypersetup{
    citecolor=freakishgreen
}

\begin{document}


\title{One Bad Apple Spoils the Bunch:\\Transaction DoS in MimbleWimble Blockchains}

\author{\IEEEauthorblockN{Seyed Ali Tabatabaee, Charlene Nicer, Ivan Beschastnikh, Chen Feng}
\IEEEauthorblockA{University of British Columbia, Canada}}

\maketitle

\begin{abstract}
As adoption of blockchain-based systems grows, more attention is being given to privacy of these systems. Early systems like BitCoin provided few privacy features. As a result, systems with strong privacy guarantees, including Monero, Zcash, and MimbleWimble have been developed.
%
Compared to BitCoin, these cryptocurrencies are much less understood.
In this paper, we focus on MimbleWimble, which uses the Dandelion++ protocol for private transaction relay and transaction aggregation to provide transaction content privacy. We find that in combination these two features make MimbleWimble susceptible to a new type of denial-of-service attacks. We design, prototype, and evaluate this attack on the Beam network using a private test network and a network simulator. We find that by controlling only $10\%$ of the network nodes, the adversary can prevent over $45\%$ of all transactions from ending up in the blockchain. We also discuss several potential approaches for mitigating this attack.

\end{abstract}


\section{Introduction}
\label{ch:introduction}


Like more established financial systems, blockchain-based digital currencies are concerned with privacy~\cite{androulaki2013evaluating}.
For example, in BitCoin~\cite{nakamoto2008bitcoin}, the transaction amounts and addresses of inputs and outputs are publicly visible. Indeed, previous research has shown that valuable information can be extracted from the resulting transaction graph, including linking of users across transactions~\cite{meiklejohn2013fistful, ober2013structure, ron2013quantitative}. In response, several protocols that offer enhanced privacy, such as Monero~\cite{cryptoeprint2015monero}, Zcash~\cite{hopwood2016zcash}, and MimbleWimble~\cite{poelstra2016mimblewimble}, have been proposed.

There are different types of privacy guarantees that users of blockchain-based networks care about and that these networks provide. One type of privacy is \emph{transaction source privacy}.
This privacy aims to hide the source of the transaction in the system~\cite{rohrer2020counting}.
To improve transaction source privacy, the Dandelion family of protocols have been proposed~\cite{bojja2017dandelion,fanti2018dandelion++}. These protocols constrain the number of neighbors that a node will send its transaction during transaction relay: a transaction will first be relayed through a \emph{stem path}, where each node passes the transaction only to one of its neighbors. This way, just one node in the network will receive the transaction directly from the transaction source and most nodes will receive the transaction once it has passed through several nodes, thereby improving source privacy.



Another important type of privacy is \emph{transaction content privacy}. This privacy aims to hide transaction content. Protocols such as Monero~\cite{cryptoeprint2015monero},
Zcash~\cite{hopwood2016zcash}, and MimbleWimble~\cite{poelstra2016mimblewimble} have introduced various techniques to enhance the content privacy of transactions. This privacy may be achieved with encryption and aggregation approaches that combine several transactions and make it difficult to reconstruct the exact set of transactions~\cite{mitani2020anonymous}.
MimbleWimble (MW) uses confidential transactions~\cite{confidentialtxs} to encrypt the amounts in transactions. And, somewhat similar to CoinJoin~\cite{maxwell2013coinjoin}, MW allows for the aggregation of several transactions into one transaction to enhance content privacy. Grin~\cite{grinnn} and Beam~\cite{beammm} are the two major implementations of MW. Both of these cryptocurrency protocols additionally use Dandelion++ for transaction relay to improve source privacy.

There are few blockchain systems that provide both source and content privacy. The MW family of blockchains is one such example. Moreover, the transaction cut-through mechanism used in MW allows for the deletion of spent outputs and thus a more compact blockchain size. Hence, MW blockchains achieve great scalability which makes them stand out from other privacy-preserving blockchain systems. However, considering the strong guarantees offered by MW blockchains, these systems have received surprisingly little research attention.

We study the MimbleWimble blockchain design in Beam from a network-level perspective, specifically focusing on its transaction relay protocol. The network-level security of blockchains has been a topic of extensive research. Multiple network-related attacks, such as eclipse attacks~\cite{heilman2015eclipse}, deanonymization attacks~\cite{biryukov2014deanonymisation, fanti2017deanonymization}, and transaction malleability attacks~\cite{decker2014bitcoin, andrychowicz2015malleability}, have been proposed against the existing blockchain systems. However, unlike this previous work, we rely on the vulnerability of the MW aggregation protocol for our proposed attack.

More specifically, the contribution of our work is the design of a new type of attack against MW blockchains that we call \textit{transaction denial of service with aggregations}. This attack targets the transaction throughput of the network. In this attack, malicious nodes on stem paths aggregate incoming transactions with a newly generated transaction that has not yet been mined into a block. As a result, at the cost of one transaction fee, an attacker can prevent all aggregations from ending up in a block. We prototyped this attack on the Beam network and found that a rogue node performing the attack can prevent $100\%$ of the incoming transactions in the stem phase from ending up in the blockchain. We demonstrated that if $10\%$ of the nodes in the network are malicious, the adversary can prevent more than $45\%$ of all transactions from ending up in the blockchain.
\section{Background}
\label{ch:background}


In this section, we provide the relevant background on Dandelion++ and MimbleWimble.

\subsection{Dandelion++ overview}
\label{sc:d++}

\begin{figure}[ht]
    \centering
    \includegraphics[width=.8\columnwidth]{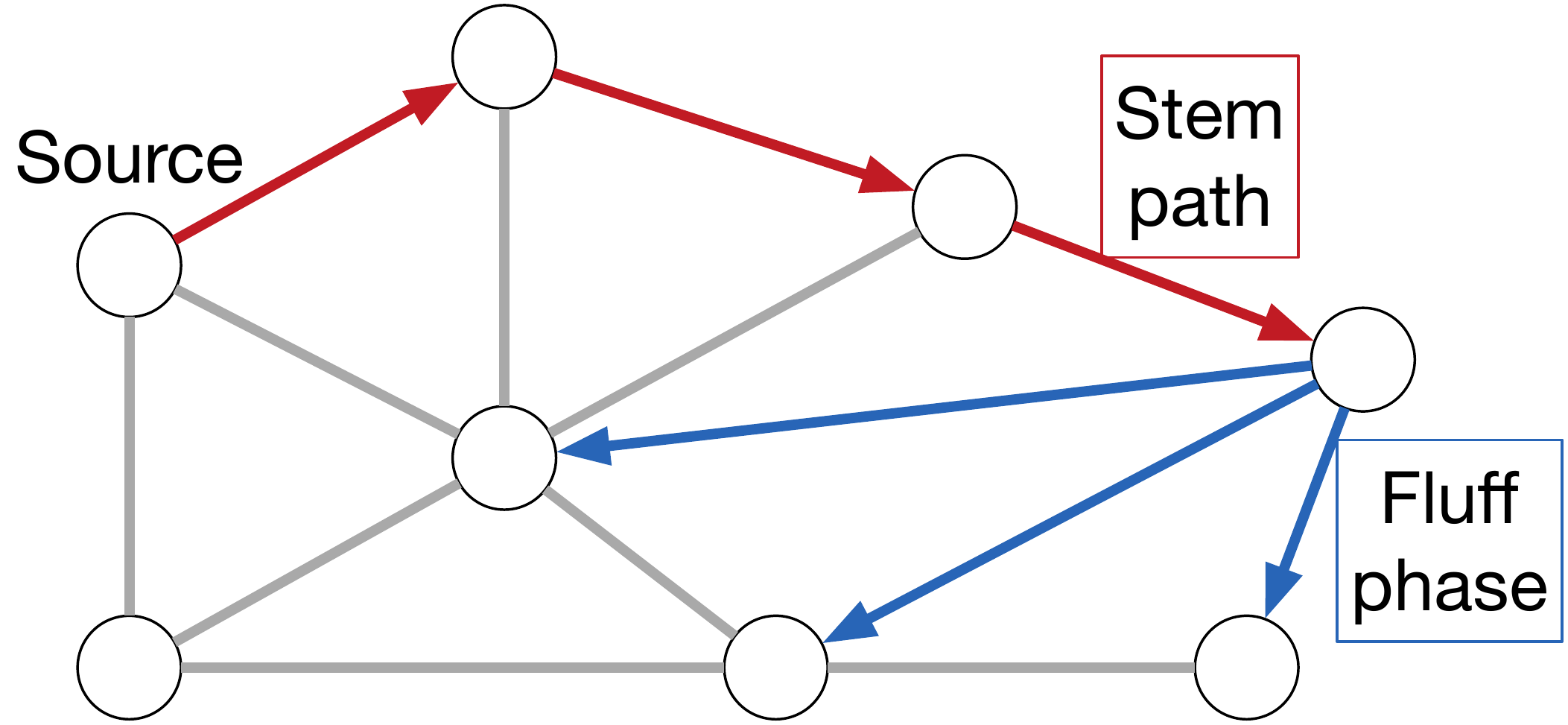}
    \caption{The two phases of transaction relay in Dandelion++. A transaction originating at node $Source$ is first relayed through a stem path. Then, the fluff phase begins and each node that receives the transaction sends it to all of its neighbors.}
    \label{fig:dandelion}
\end{figure}

Dandelion++ is a transaction relay protocol based on a preceding proposal called Dandelion. The two protocols have similar goals but subtle differences in implementation choices. To improve the source privacy of transactions, Dandelion++ constrains the number of neighbors that a node will send a transaction at the beginning of the transaction relay. With Dandelion++, transactions are relayed in two phases. First, in the stem phase, when a node receives a transaction it passes the transaction to just one other node. Then, in the fluff phase, a node forwards a received transaction to all of its neighbors except the one that initially sent it the transaction. Figure~\ref{fig:dandelion} illustrates these two phases of transaction relay in Dandelion++.

Compared to the broadcast-based transaction dissemination protocol used in BitCoin, in Dandelion++ adversarial nodes have a less chance of receiving a transaction directly from the transaction source so it is more difficult for the adversary to localize the transaction source node. The probability of transitioning to the fluff phase in each step of the stem phase is a parameter in the protocol. The lower this probability is, the longer the average length of stem paths. Longer stem paths improve transaction source privacy, but increase latency. To mitigate black-hole attacks where the adversarial nodes decide not to forward an incoming stem transactions, Dandelion++ incorporates a fail-safe mechanism. Each node along the stem path of a transaction would fluff the transaction on its own if it does not receive the fluff version of the transaction within a time period. For this, nodes along the stem path create independent random timers for the transaction.

\subsection{MimbleWimble (MW) overview}
\label{sc:mw}

MimbleWimble is a cryptocurrency protocol that uses encryption and aggregation to enhance the content privacy of transactions. Compared to other cryptocurrency protocols like Bitcoin, MimbleWimble has the following important advantages:

\begin{itemize}
  \item Input and output amounts in a transaction are encrypted.
  \item Aggregation of transactions makes it difficult to link the inputs and outputs.
  \item The size of the blockchain is reduced through the deletion of spent outputs (the cut-through mechanism).
\end{itemize}

MimbleWimble uses confidential transactions to encrypt amounts. The commitments of inputs and outputs are put into transactions and kept on the blockchain. Each commitment is in the form of

\[C = r \cdot G + v \cdot H\]
where $C$ is a Pedersen commitment, $v$ is the amount, $r$ is a secret random blinding key which should be known only to the owner, and $G$ and $H$ are fixed Elliptic Curve Cryptography (ECC) group generators known to all. A range-proof is attached to each output commitment which proves that its amount is valid. The $r$ value in commitments with explicit amounts, such as transaction fees and block rewards, is zero. The owner of a set of outputs is one who knows the sum of their $r$ values. By knowing the sum of $r$ values for a set of outputs, one can create a valid transaction that spends those outputs. For a transaction to be valid, the commitments in that transaction should sum to zero and the range-proofs for the output commitments should be valid.

To prevent the sender of inputs in a transaction from spending the outputs, the sum of $r$ values for the outputs should differ from the sum of $r$ values for the inputs. Therefore, the commitments of inputs and outputs in each transaction should sum to a non-zero value $k \cdot G$ (kernel) where $k$ is chosen by the recipient. A kernel is a non-spendable commitment with zero amount. A transaction is allowed to have more than one kernel. Hence, the sum of commitments in a transaction is

\[\sum_{C_i \in inputs}{C_i} + \sum_{C_o \in outputs}{C_o} + \sum_{C_k \in kernels}{C_k} = 0.\]

Aggregation of transactions makes it difficult to link the inputs and outputs. Since the sum of commitments in each valid transaction is zero, the total sum of commitments for multiple transactions is still zero. Therefore, the aggregation of multiple valid transactions is a valid transaction. 
Because each block consists of some valid transactions, a block can be interpreted as a single aggregated transaction.

\begin{figure}[ht]
    \centering
    \includegraphics[width=.95\columnwidth]{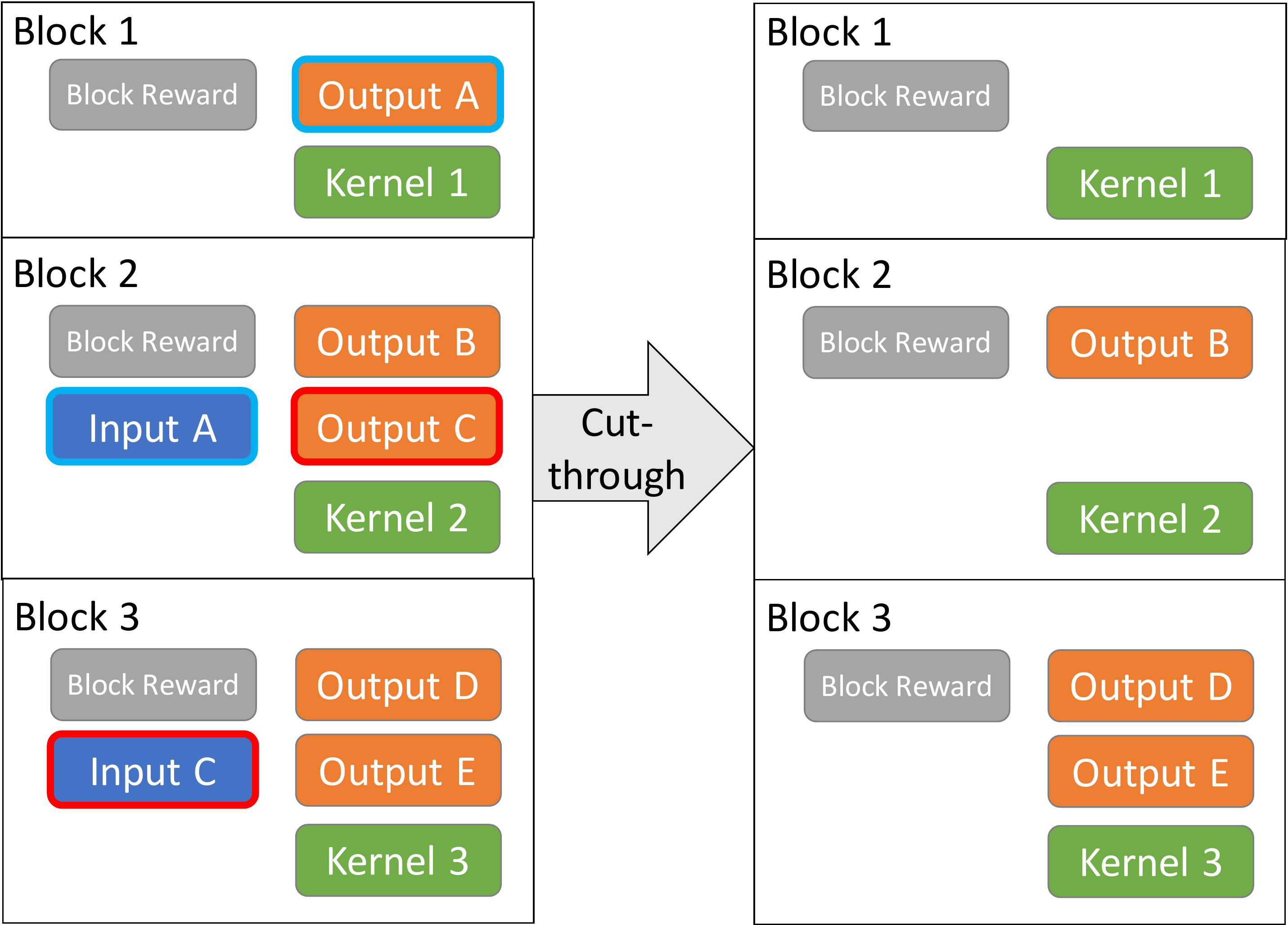}
    \caption{Deletion of spent outputs and their corresponding inputs from the blockchain in MimbleWimble.}
    \label{fig:txdel}
\end{figure}

Transaction cut-through is one of the most significant features of MimbleWimble. In MimbleWimble, it is possible to safely remove a spent output and its corresponding input from the blockchain (Figure~\ref{fig:txdel}). 
The sum of all commitments in each block is zero. Hence, the sum of all commitments in the blockchain is also zero. An output of a transaction can be spent in the succeeding blocks and appear as an input of another transaction. The sum of commitments for this pair of input and output is zero. Consequently, if both commitments are removed from the blockchain, the sum of all commitments in the blockchain remains zero. Using this technique, the size of the blockchain can be substantially reduced. The only elements that remain in the blockchain are the explicit amounts for block rewards, kernels for all transactions, and unspent outputs along with their range-proofs and Merkle proofs.

Although the original proposal did not specify a transaction relay protocol for MimbleWimble, the two major implementations of this protocol, Grin~\cite{grinnn} and Beam~\cite{beammm}, have incorporated Dandelion++, where transactions are aggregated in the stem phase. Using this approach, not only do these cryptocurrencies attempt to improve the transaction source privacy, but also they try to make it difficult to link the inputs and outputs of transactions by first relaying them through stem paths and reducing the number of network nodes that observe them before aggregation.

Bulletproofs~\cite{bunz2018bulletproofs}, which are short proofs for confidential transactions, have also been proposed to improve on the original range-proofs in MimbleWimble. Bulletproofs have been adopted by both Grin and Beam implementations. Other research projects have provided a provable-security analysis for MimbleWimble~\cite{fuchsbauer2019aggregate} as a step toward a formalization of the MimbleWimble protocol and a verification of its implementations~\cite{betarte2020towards, silveira2021formal}.

In this project, we focus on the vulnerabilities of transaction relay in the implementations of MimbleWimble.
\section{Transaction Relay in Beam}
\label{ch:beam}

 Here, we describe Beam's transaction relay protocol. We provide an overview of the life cycle of a transaction from when it is received by a node to when it is forwarded to the peers of the node. For that purpose, we use pseudocode that we have written based on Beam's source code. All the pseudocode presented here has been obtained from the node/node.cpp file in the ``mainnet'' branch of Beam's GitHub repository as of February 2021\footnote{\url{https://github.com/BeamMW/beam/commit/ade19e1f8b1a702ad81d81092ba6a8f6561513ed}}.

There are six important functions that every Beam node uses to manage and forward incoming transactions. These are also the functions that we need to modify in the source code to implement our proposed attack. Table~\ref{tab:beampars} presents the values for the Beam network parameters that are used in the functions that we describe.

\begin{table} 
\caption{The Beam network parameters.}
\centering
\begin{tabular}{rl}
 \hline
 \textbf{Name} & \textbf{Value}\\
\hline
 FluffProbability  & 0.1 \\
\hline
 TimeoutMin  & 20s \\
\hline
 TimeoutMax  & 50s \\
\hline
 AggregationTime  & 10s \\
\hline
 OutputsMin  & 5 \\
\hline
 OutputsMax  & 40 \\
\hline
\end{tabular}
\label{tab:beampars}
\end{table}

When a new transaction is received, the function \emph{OnTransaction} (Algorithm~\ref{alg:OnTransaction}) will be called. This function calls either \emph{OnTransactionStem} (Algorithm~\ref{alg:OnTransactionStem}) or \emph{OnTransactionFluff} (Algorithm~\ref{alg:OnTransactionFluff}) based on the type of the incoming transaction.

\begin{algorithm} [ht] \footnotesize
    \caption{OnTransaction}\label{alg:OnTransaction}
    \begin{algorithmic}[1]
        \Function{OnTransaction}{Transaction $tx$}
            \If{$tx$ is stem}
                \State OnTransactionStem($tx$)
            \Else
                \State OnTransactionFluff($tx$)
            \EndIf
        \EndFunction
    \end{algorithmic}
\end{algorithm}

If the incoming transaction is a stem transaction, then the function \emph{OnTransactionStem} (Algorithm~\ref{alg:OnTransactionStem}) will be called. This function compares the new stem transaction to transactions in the node's stempool (data structure containing valid stem transactions that have not been fluffed) and checks the validity of the new transaction. If the new stem transaction is accepted, then the stempool will be updated and the new transaction will also be added to it. Eventually, if the number of outputs in the transaction is greater than or equal to \emph{OutputsMax}, then the transaction does not need to be aggregated any further; hence, the function \emph{OnTransactionAggregated} (Algorithm~\ref{alg:OnTransactionAggregated}) will be called. Otherwise, \emph{PerformAggregation} (Algorithm~\ref{alg:PerformAggregation}) will be called.

\begin{algorithm} [ht] \footnotesize
    \caption{OnTransactionStem}\label{alg:OnTransactionStem}
    \begin{algorithmic}[1]
        \Function{OnTransactionStem}{Transaction $tx$}
            \For{each Kernel $k$ in $tx$} \Comment{at most one Tx in stempool has $k$}
                \State Find Transaction $q$ in stempool that contains $k$ \Comment{if it exists}
                \State // continue to the next iteration if such $q$ does not exist
                \If{$tx$ does not cover $q$} \Comment{$tx$ covers $q$ if it has all Kernels of $q$}
                    \State Drop $tx$
                    \State \textbf{return} \Comment{error code will be returned to sender}
                \EndIf
                \If{$q$ covers $tx$} \Comment{it means $tx$ and $q$ are the same}
                    \If{$q$ is still aggregating} \Comment{should not normally happen}
                        \State Drop $tx$
                        \State \textbf{return} \Comment{with 'accept' error code}
                    \Else
                        \State \textbf{break}
                    \EndIf
                \EndIf
                \State // if $tx$ covers $q$ but $q$ does not cover $tx$
                \State Validate($tx$) \Comment{if not done before}
                \State Drop $q$ from stempool
            \EndFor
            \State Validate($tx$) \Comment{if not done before}
            \State // by this point, the given stem-tx is accepted
            \State Add $tx$ to stempool \Comment{also add dummy inputs to $tx$ if necessary}
            \If{NoNeedForAggregation($tx$)} \Comment{$tx$ has at least \emph{OutputsMax} outputs}
                \State OnTransactionAggregated($tx$)
            \Else
                \State PerformAggregation($tx$)
            \EndIf
        \EndFunction
    \end{algorithmic}
\end{algorithm}

Given a stem transaction, \emph{OnTransactionAggregated} (Algorithm~\ref{alg:OnTransactionAggregated}) sends the stem transaction to a randomly chosen peer with a probability of $0.9$ or fluffs the transaction by calling \emph{OnTransactionFluff} (Algorithm~\ref{alg:OnTransactionFluff}) with a probability of $0.1$.

\begin{algorithm} [ht] \footnotesize
    \caption{OnTransactionAggregated}\label{alg:OnTransactionAggregated}
    \begin{algorithmic}[1]
        \Function{OnTransactionAggregated}{Transaction $tx$}
            \If{$RandInt(1, 10) \neq 10$}
                \State Select a random Peer $p$
                \State Send (stem) $tx$ to $p$
                \State Set timer (uniformly selected between \emph{TimeoutMin} and \emph{TimeoutMax}) on $tx$ to later check if it is fluffed or not
            \Else \Comment{\emph{FluffProbability} = $0.1$}
                \State OnTransactionFluff($tx$)
            \EndIf
        \EndFunction
    \end{algorithmic}
\end{algorithm}

\emph{PerformAggregation} (Algorithm~\ref{alg:PerformAggregation}) tries to merge a given stem transaction with other transactions in the stempool. In the end, if the number of outputs in the transaction is at least \emph{OutputsMin} (the transaction does not necessarily need more aggregation), \emph{OnTransactionAggregated} (Algorithm~\ref{alg:OnTransactionAggregated}) will be called to forward the transaction. If the transaction still needs to be aggregated, the function will set a timer ($10$s) on the transaction to bound the time that it remains in the stempool without being forwarded.

\begin{algorithm} [ht] \footnotesize
    \caption{PerformAggregation}\label{alg:PerformAggregation}
    \begin{algorithmic}[1]
        \Function{PerformAggregation}{Transaction $tx$}
            \For{each Transaction $q$ in stempool that needs to be aggregated, starting from the one with the closest profitability to $tx$, until !NeedsAggregation($tx$)} \Comment{in Beam, Transaction profitability is defined as $\frac{Transaction\ fee}{Transaction\ size}$}
                \State TryMerge($tx, q$) \Comment{merges $q$ into $tx$ if the result is valid}
            \EndFor
            \State 
            \If{$tx$ has at least \emph{OutputsMin} outputs}
                \State OnTransactionAggregated($tx$)
            \Else
                \State Set timer (\emph{AggregationTime}) on $tx$ \Comment{to later add dummy outputs and stem if not aggregated enough by then}
            \EndIf
        \EndFunction
    \end{algorithmic}
\end{algorithm}

The function \emph{OnTransactionFluff} (Algorithm~\ref{alg:OnTransactionFluff}), after making sure that a given transaction is valid, updates the stempool. The function also updates the fluffpool (data structure containing valid fluff transactions) and sends the given transaction to all of its peers except the one that initially sent the fluff transaction.

\begin{algorithm} [ht] \footnotesize
    \caption{OnTransactionFluff}\label{alg:OnTransactionFluff}
    \begin{algorithmic}[1]
        \Function{OnTransactionFluff}{Transaction $tx$}
            \If{$tx$ is in stempool}
                \State Drop $tx$ from stempool
            \EndIf
            \If{$tx$ is already in fluffpool} \Comment{we already received the fluff tx}
                \State Drop $tx$
                \State \textbf{return} \Comment{with 'accept' error code}
            \EndIf
            \State Validate($tx$) 
            \If{$tx$ was not in stempool} \Comment{when this function was called}
                \For{each Kernel $k$ in $tx$}
                    \State Find Transaction $q$ in stempool that contains $k$ \Comment{if it exists}
                    \State // continue to the next iteration if such $q$ does not exist
                    \State Drop q from stempool \label{line:attack}
                \EndFor
            \EndIf
            \While{fluffpool does not have enough capacity for $tx$}
                \State Find q, the least profitable Transaction in fluffpool
                \If{$q$ is less profitable than $tx$}
                    \State Drop $q$ from fluffpool
                \Else
                    \State Drop $tx$
                    \State \textbf{return} \Comment{with 'accept' error code}
                \EndIf
            \EndWhile
            \State Add $tx$ to fluffpool
            \State Send $tx$ to all Peers of the node \Comment{except the Peer that sent $tx$}
        \EndFunction
    \end{algorithmic}
\end{algorithm}



Finally, we explain \emph{OnTimedOut} (Algorithm~\ref{alg:OnTimedOut}). If a stem transaction is still waiting for aggregation by the expiration of the timer that was set for it in the \emph{PerformAggregation} function (Algorithm~\ref{alg:PerformAggregation}), then dummy outputs will be added to the transaction (to ensure that the transaction has at least \emph{OutputsMin} outputs and therefore it is sufficiently difficult to link its inputs and outputs) and \emph{OnTransactionAggregated} (Algorithm~\ref{alg:OnTransactionAggregated}) will be called to forward the transaction. Moreover, if the fluff version of a forwarded stem transaction is not received by the expiration of its independent random timer, then \emph{OnTransactionFluff} (Algorithm~\ref{alg:OnTransactionFluff}) will be called to fluff the transaction.

\begin{algorithm} [ht] \footnotesize
    \caption{OnTimedOut}\label{alg:OnTimedOut}
    \begin{algorithmic}[1]
        \Function{OnTimedOut}{Transaction $tx$}
            \If{$tx$ is still aggregating}
                \State Add dummy outputs to $tx$ so that it has at least \emph{OutputsMin} outputs
                \State OnTransactionAggregated($tx$)
            \Else \Comment{fluff timed-out, emergency fluff}
                \State OnTransactionFluff($tx$)
            \EndIf
        \EndFunction
    \end{algorithmic}
\end{algorithm}

\section{Threat Model}
\label{ch:model}






The participating nodes form a peer-to-peer network. The adversary in our model can create nodes and connect to other nodes in the network. The adversarial nodes can connect to more nodes than what the protocol suggests. The adversary needs to know the addresses of other nodes before connecting to them. Nevertheless, the adversary cannot impose a connection on any other node if the other node does not want to connect to it. By increasing the number of adversarial nodes in the network or the number of connections from adversarial nodes to honest nodes, the adversary will be incident on more relay paths and therefore can attack the transaction relay paths more effectively.

We assume that instead of targeting specific nodes or users, the adversary is interested in mass attacks on the honest portion of the network. Nonetheless, selective attacking could help to hide the position of the adversary in the network. Adversarial nodes can store the information that they receive about the network and the transactions. They can analyze the stored information and adjust their decisions. The adversary can deviate from the relay policy of the network and disregard the relay phase of transactions. Also, the adversary can generate new valid transactions and pay for their transaction fees. The adversarial nodes can aggregate different valid transactions that they previously received or generated.

The adversary in our model is only interested in attacking the transaction relay network and does not influence the block generation process. Therefore, we do not assume any mining power for the adversary. Generally, the adversary is unaware of the exact topology of the network and the connections between pairs of honest nodes. We assume that the adversary cannot decrypt commitments in transactions to learn their amounts or secret blinding keys. Furthermore, the adversary cannot spend the outputs that are owned by others or trick honest nodes into accepting invalid transactions.

\section{Approach}
\label{ch:approach}

To improve content privacy, MimbleWimble allows for the aggregation of transactions. However, the adversary can exploit this feature to launch a denial of service attack. Among different aggregations that have a transaction in common, at most one can end up in the blockchain. Therefore, by aggregating different incoming transactions with a newly generated transaction, the adversary can perform a denial of service attack on the incoming transactions.

Let $T_A$ be a new stem transaction received by an adversarial node. To execute the attack, instead of normally aggregating and relaying the stem transaction, the adversarial node generates a new transaction $T_B$. Then, the adversarial node aggregates the two transactions into $T_A + T_B$ and fluffs both $T_A + T_B$ and $T_B$. Since $T_A$ is fluffed as a part of an aggregated transaction, the other nodes in the stem path of $T_A$ will not separately fluff $T_A$. Nevertheless, only one transaction between $T_A + T_B$ and $T_B$ can end up in the blockchain. Hence, if the adversarial node creates $T_B$ in a way that miners prioritize it over $T_A + T_B$ (for this to happen, the profitability for $T_B$ should be higher than the profitability for $T_A + T_B$), then $T_A$ will not end up in the blockchain. In this case, the wallet that initially created $T_A$ needs to resend $T_A$ to the network. The cost of this denial of service attack for the adversary is the transaction fee of $T_B$.

To validate the feasibility of this attack, in our attack, a rogue node will aggregate the incoming stem transactions with new transactions that have not been mined into any block and fluff the resulting aggregations and the newly generated transactions. 
We will then measure and compare the block mining time for the transactions that were aggregated in this way by the adversary and for the normally relayed transactions. 
\section{Implementation}
\label{ch:implementation}

In this section, we explain our implementation of the Beam network simulator, the proposed attack, and the Beam private test network.

We have provided a simulation of the Beam network to estimate the percentage of stem paths that the adversary will be incident on. The inputs of this simulation are parameters such as the number of nodes, the percentage of malicious nodes, the expected degree of each node, and the probability of transitioning to the fluff phase in each step of the stem phase. Based on these parameters, the program creates a pseudorandom graph representing the network. The connections of each node are uniformly selected among all other nodes without replacement. The program then tests 1~million pseudorandom stem paths for the estimation. Each tested stem path starts from a source uniformly selected from the set of all nodes and each node in the stem path selects the next node uniformly from the set of its neighbors to forward the stem transaction. We implemented this network simulator in approximately $200$ lines of C++.

To implement our proposed attack, we modified the source code of Beam in the "testnet" branch\footnote{\url{https://github.com/BeamMW/beam/tree/testnet}}. Most of the changes were applied to the files in the "node" directory and especially the functions described in Section~\ref{ch:beam}. We modified approximately $500$ lines of C++ code to implement our attack.

To validate our proposed attack, we have implemented a private test network. The network consists of some normal Beam nodes and some malicious nodes that run with our modifications. This private network has the following properties:

\begin{enumerate}

\item \textbf{Number of nodes:} After consulting with the Beam developer community, considering the requirements of our evaluations, and also taking into account the resources available, we decided to have $100$ nodes in our private network.

\item \textbf{Number of bootstrapping nodes:} The Beam main network includes multiple bootstrapping nodes in different geographical locations\footnote{\url{https://beam.mw/downloads/mainnet-mac}}. When a new node joins the Beam network, it usually connects to bootstrapping nodes at the beginning. Hence, bootstrapping nodes have more connections and receive more stem transactions compared to normal nodes. In our private network, there are $10$ bootstrapping nodes (out of a total of $100$ nodes) and normal nodes first connect to them upon joining the network.

\item \textbf{Number of adversarial nodes:} This is a configurable parameter of our private test network. For different attacks, we might need different numbers of adversarial nodes.

\item \textbf{Versions of nodes:} For the honest nodes, we use the latest stable version of the "testnet" branch\footnote{\url{https://github.com/BeamMW/beam/commit/cfe091468fbcfcd2092352c22a18099bf9d017f0}}. For the adversarial nodes, we modify this source code to implement each of our proposed attacks.

\item \textbf{Number of connections per node:} We do not change the algorithm for finding new connections and we maintain all the node policies from the Beam main network in the test network.

\item \textbf{Probability of transitioning to the fluff phase:} Similar to the policy of the Beam main network and test network, we set the probability of transitioning to the fluff phase in each step of the stem phase in our private network to $0.1$.

\item \textbf{Mining:} 
In our private test network, similar to the Beam main network and test network, a new block is added to the blockchain every minute, on average.
The Beam main network uses a Proof of Work (PoW) scheme to grow the blockchain. Instead, in our private test network, we use a fake mining scheme to avoid wasting our resources on the expensive process of PoW mining. In the fake mining scheme, nodes do not compete with each other over mining new blocks. Fake mining is adequate since our attack focuses on transaction relay and does not assume malicious miners.

\item \textbf{Transaction generation rate:} This is a configurable parameter of our private network. For most of our experiments, we want to set the transaction generation rate in a way that the frequency of stem transactions received in the private network nodes reflects the frequency in the Beam main network. Nevertheless, we also want to be able to vary the transaction generation rate to observe its effect on aggregations.

\item \textbf{Number of wallets assigned to each node:} We assign one wallet to each node of the network because we want newly generated transactions to be relayed from each node.

\end{enumerate}

We deployed our private test network
across Azure virtual machines running Ubuntu Server 18.04 LTS Gen 1. We launched 100 virtual nodes in two geographically separated VMs located in the Eastern United States and South East Asia with each server containing $50$ virtual nodes.
The latency between the virtual nodes in our setup is simulated by adding a pseudorandom delay from a normal distribution with a mean of $100$ms and a standard deviation of $20$ms to each message using the NetEm network emulator. The topology of the network is controlled by the peer parameter in the command line interface of the Beam node. Transactions are generated using the Beam wallet Node.js API.
\section{Evaluation}
\label{ch:evaluation}

In this section, we evaluate our proposed attack on the implementation of MimbleWimble in Beam. In particular, we focus on the following question:

\begin{itemize}

\item How does maliciously aggregating a transaction with other transactions increase the transaction processing time from the user's perspective?

\end{itemize}

To answer the question above, we need to determine the proportion of transactions attacked by the adversarial nodes and the impact of our proposed attack on a targeted transaction. We use our network simulator to estimate the proportion of transactions that the adversary can attack, given the percentage of adversarial nodes and other network parameters. To determine the impact of our proposed attack on a targeted transaction, we run a rogue node in our test network and perform the attack.

\textbf{Simulation results.}
The adversary can perform the attack on a transaction if and only if an adversarial node is on the stem path of that transaction. Therefore, to estimate the proportion of transactions that the adversary can attack, we have conducted several experiments with our network simulator to estimate the percentage of stem paths that the adversary will be incident on (also referred to as infected stem paths), given the network parameters.

Table~\ref{tab:simpars} presents the default values that we have used for the parameters of the network simulator.
We have set the probability of transitioning to the fluff phase to $0.1$ (similar to the policy of the Beam main network), the number of nodes to $1000$, and the expected degree of each node to $10$.
We use $10$ based on a measurement study: we recorded the number of connections for a Beam main network node and a Beam test network node, deployed on our University servers, every six hours from April 21 2021 to April 23 2021. We observed that the recorded numbers were between $11$ and $14$ for the main network node and between $9$ and $11$ for the test network node. Therefore, setting the default value for the expected degree of each node to $10$ is reasonable.

In each experiment, we have varied the value of one network parameter while maintaining the default values for other parameters. Hence, we have observed the effect of each network parameter on the percentage of stem paths that pass through the adversary.

\begin{table} 
\caption{The default values for the parameters of the network simulator.}
\centering
\begin{tabular} {rl}
 \hline
 \textbf{Name} & \textbf{Value}\\
\hline
 Percentage of Malicious Nodes  & 10\% \\
\hline
 Probability of Transitioning to the Fluff Phase  & 0.1 \\
\hline
 Number of Nodes  & 1000 \\
\hline
 Expected Degree of Each Node  & 10 \\
\hline
 Number of Bootstrapping Nodes  & 0 \\
\hline
\end{tabular}
\label{tab:simpars}
\end{table}

\begin{figure}[ht]
\begin{subfigure}{.24\textwidth}
    \centering
    \includegraphics[width=1\linewidth]{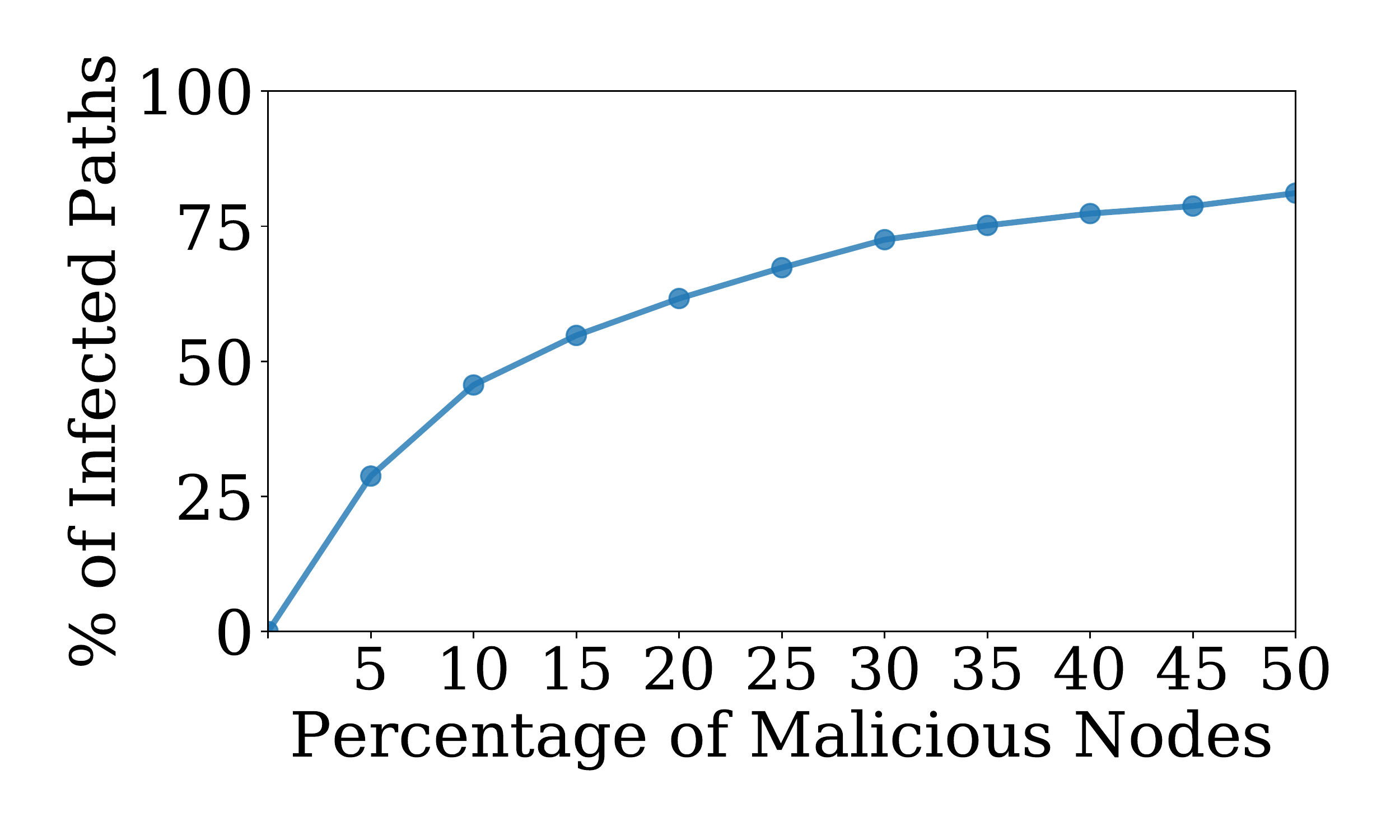}
    \caption{}
    \label{fig:mal-infected}
\end{subfigure}
\begin{subfigure}{.24\textwidth}
    \centering
    \includegraphics[width=1\linewidth]{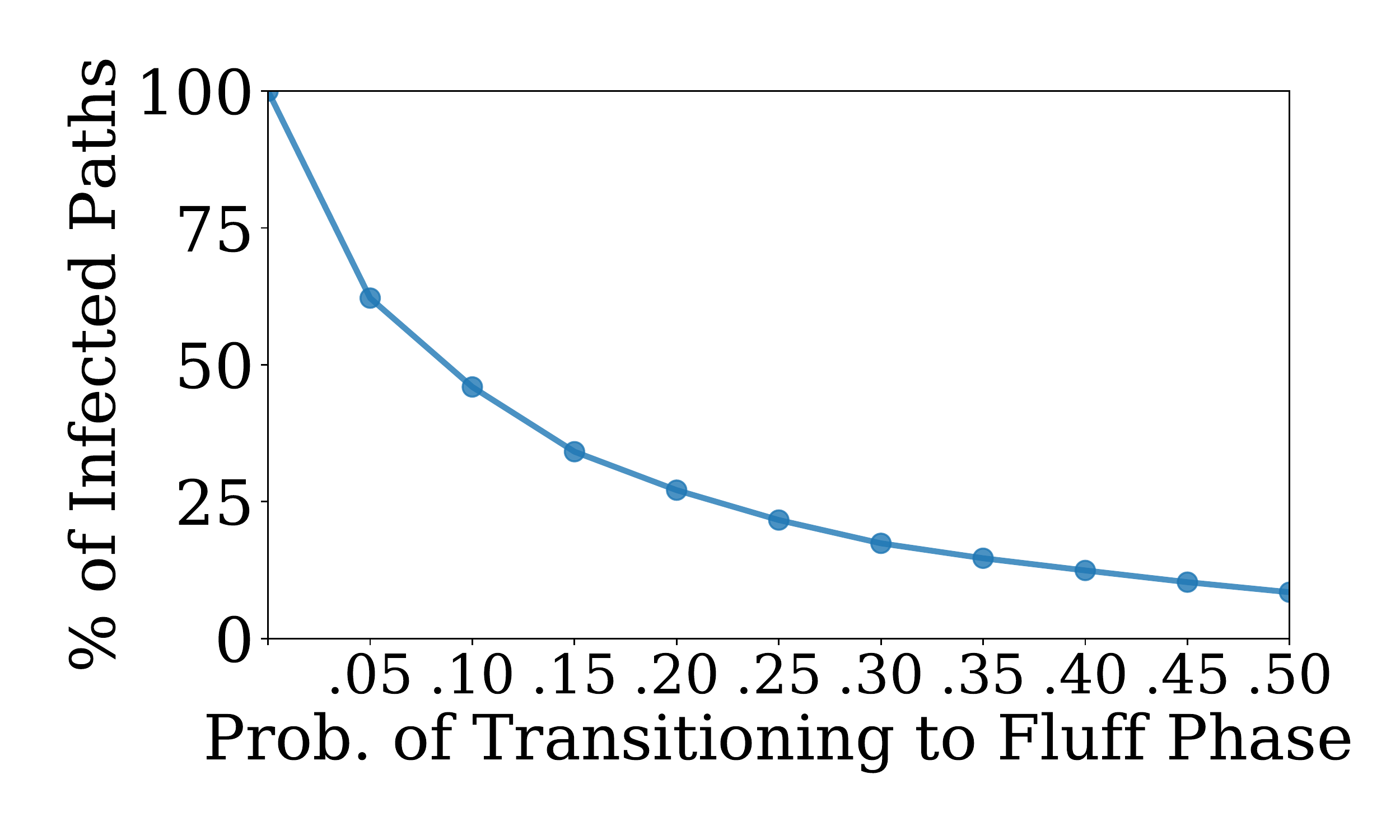}
    \caption{}
    \label{fig:fluff-infected}
\end{subfigure}
\begin{subfigure}{.24\textwidth}
    \centering
    \includegraphics[width=1\linewidth]{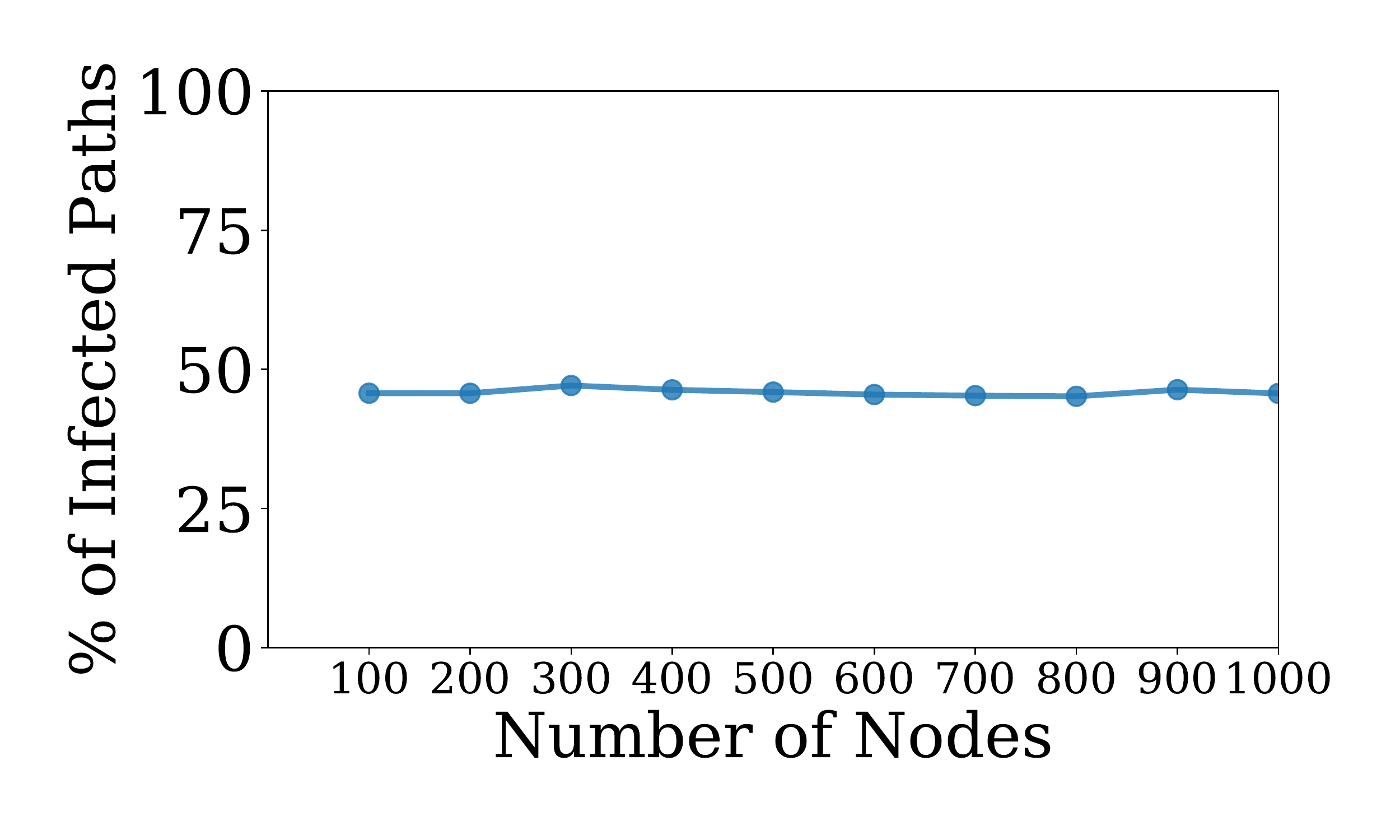}
    \caption{}
    \label{fig:nodes-infected}
\end{subfigure}
\begin{subfigure}{.24\textwidth}
    \centering
    \includegraphics[width=1\linewidth]{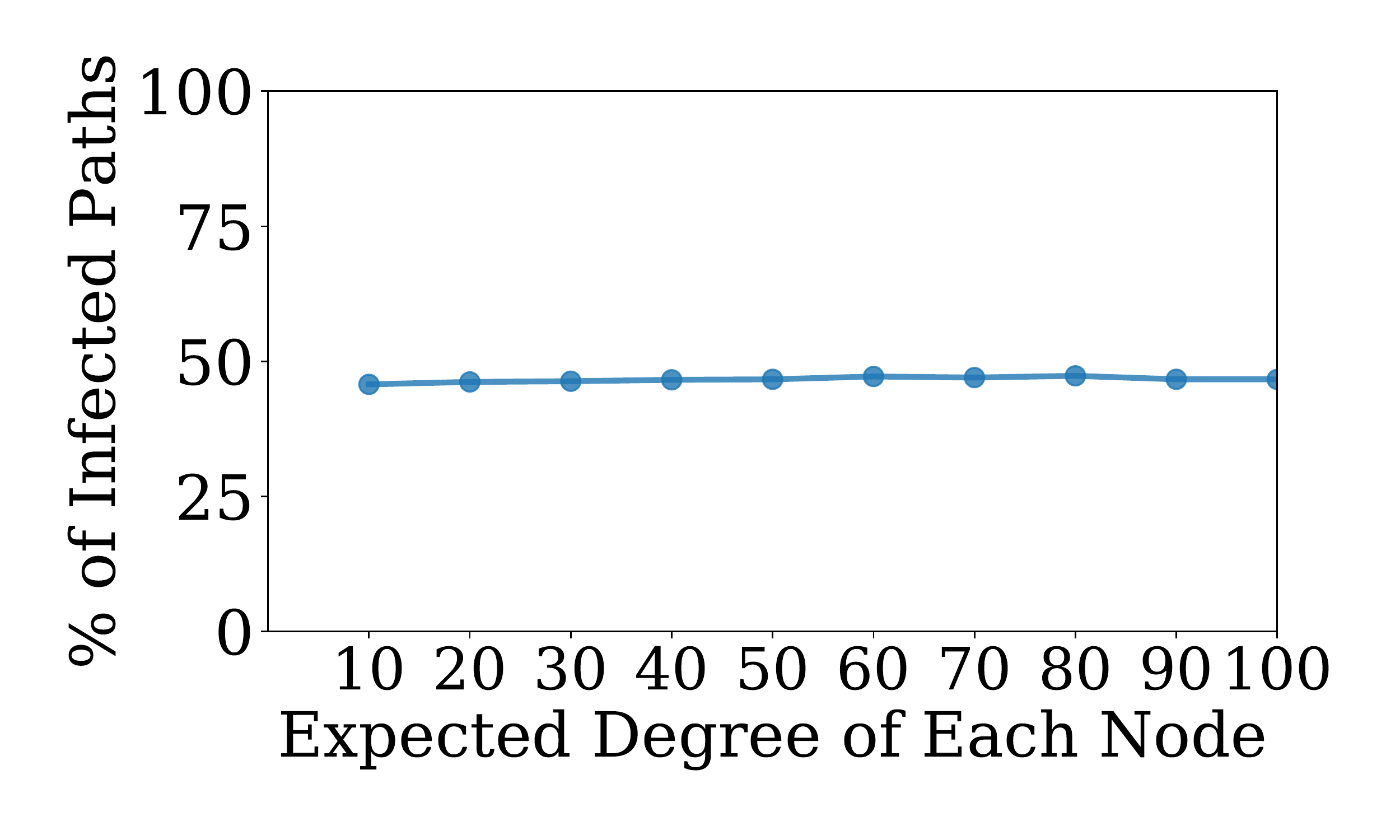}
    \caption{}
    \label{fig:degree-infected}
\end{subfigure}
\begin{subfigure}{.24\textwidth}
    \centering
    \includegraphics[width=1\linewidth]{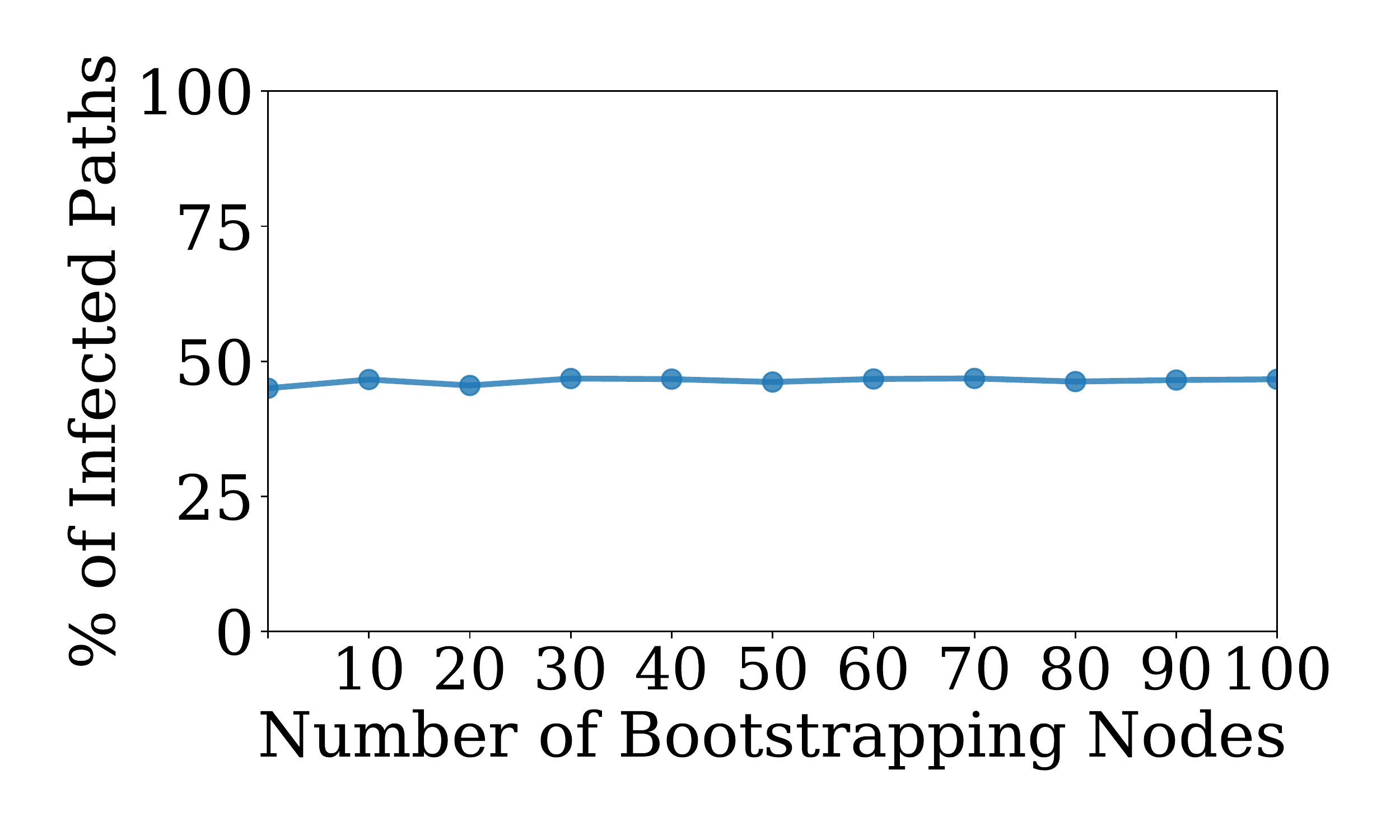}
    \caption{}
    \label{fig:bootstr-infected}
\end{subfigure}
\caption{The percentage of stem paths that the adversary will be incident on, also referred to as infected paths, with (a) varying percentages of malicious nodes, (b) varying probabilities of transitioning to the fluff phase in each step of the stem phase, (c) varying numbers of nodes, (d) varying expected degrees of nodes, and (e) varying numbers of bootstrapping nodes.
}
\label{fig:simul}
\end{figure}

Figure~\ref{fig:simul} shows the results of our experiments with the network simulator. We observe that by increasing the percentage of malicious nodes in the network, the adversary will be incident on more stem paths and therefore can attack more transactions (Figure~\ref{fig:mal-infected}). We also observe that increasing the probability of transitioning to the fluff phase in each step of the stem phase (and consequently decreasing the average length of stem paths) decreases the percentage of stem paths that pass through the adversary (Figure~\ref{fig:fluff-infected}). Nevertheless, varying the values of other network parameters, such as the number of nodes (Figure~\ref{fig:nodes-infected}), the expected degree of each node (Figure~\ref{fig:degree-infected}), and the number of bootstrapping nodes (Figure~\ref{fig:bootstr-infected}), does not particularly affect the percentage of stem paths that pass through the adversary. 
The probability that a stem transaction passes through the adversary depends on the probability that the transaction arrives at an adversarial node in each step of the stem phase and also on the number of steps. The percentage of adversarial nodes in the network determines the probability that a transaction arrives at such a node in each step of the stem phase. Furthermore, the probability of transitioning to the fluff phase in each step of the stem phase determines the average number of steps. These are the reasons why the percentage of stem paths that the adversary will be incident on depends on the percentage of adversarial nodes in the network and the probability of transitioning to the fluff phase, but not the other parameters.

Let us consider the cases where the probability of transitioning to the fluff phase in each step of the stem phase is set to $0.1$ (similar to the Beam main network). We note that if $10\%$ of the nodes are malicious, the adversary will be incident on the stem paths of more than $45\%$ of all transactions in the network and hence can attack those transactions. Increasing the percentage of malicious nodes to $30\%$ increases the percentage of transactions that the adversary can attack to over $70\%$.

\textbf{Testnet results.}
%
To measure the impact of our attack, we added a rogue node to a private test network. Our rogue node performed the attack on $300$ incoming stem transactions. For each attacked transaction $T_A$, our rogue node generated a new transaction $T_B$ with higher profitability compared to $T_A$ and fluffed $T_A + T_B$ and $T_B$. For each attacked transaction $T_A$ and its corresponding adversarial transaction $T_B$, we observed whether $T_A + T_B$ or $T_B$ ended up in the blockchain.

Our rogue node successfully prevented $100\%$ of the attacked transactions from ending up in the blockchain. In fact, for each attacked transaction $T_A$ and its corresponding adversarial transaction $T_B$, $T_A + T_B$ had lower profitability compared to $T_B$ and hence $T_B$ ended up in the blockchain. Therefore, if $10\%$ of the nodes are malicious, the adversary can attack more than $45\%$ of all transactions and prevent them from ending up in the blockchain.

We have also measured the latency of $300$ normally relayed stem transactions and compared this latency against the latency of transactions that the adversary generated to perform the denial of service attack. We measured the latency of each normally relayed transaction by calculating the difference between the time that our node received that transaction in the stem phase and the time that the transaction was recorded in the blockchain (obtained from the block timestamp). We measured the latency of each adversarial transaction by determining the difference between the time that our rogue node generated that transaction and the time that the transaction was recorded in the blockchain. 

\begin{figure}[ht]
    \centering
    \includegraphics[width=\columnwidth]{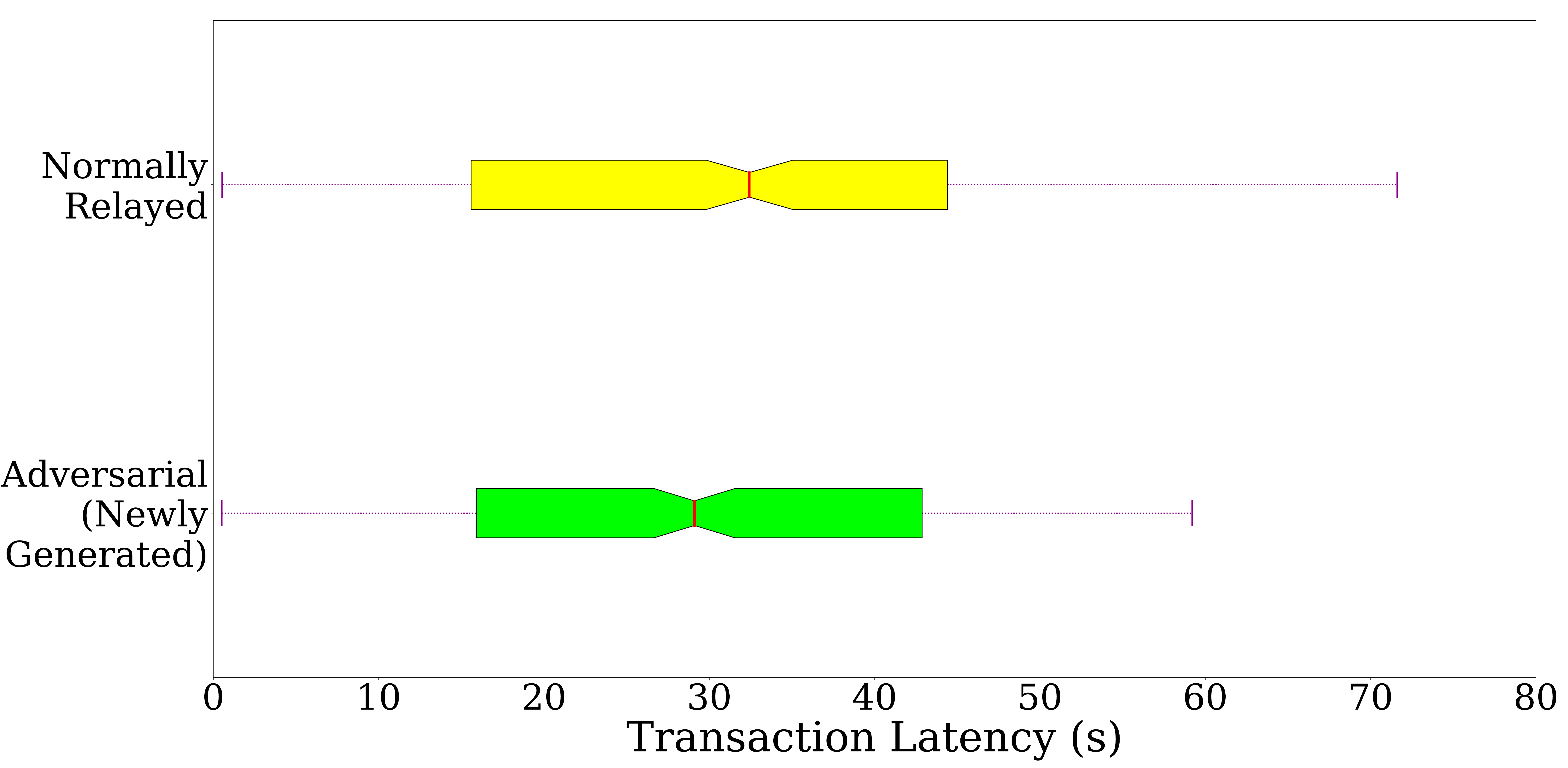}
    \caption{Latency distribution comparison of the normally relayed transactions and the transactions that the adversary generated to perform the DoS attack with aggregations.
    }
    \label{fig:aggrattack}
\end{figure}

Figure~\ref{fig:aggrattack} shows the latency results. The average latency for the adversarial transactions in the attack is $29$s and it is slightly lower than the average latency for normally relayed transactions, which is $31$s. That is because the adversary immediately fluffs the newly generated transactions to perform the attack.

\section{Discussion}
\label{ch:discussion}

\textbf{Cost of attack.}
The cost of the attack for the adversary are the transaction fees of newly generated transactions. The adversary can reduce the cost for a newly generated transaction by reducing its size while choosing a sufficient transaction fee so that the profitability (defined as $\frac{Transaction\ fee}{Transaction\ size}$) of the new transaction is higher than the profitability of the incoming stem transaction that it is aggregated with. Moreover, if the adversary modifies each of its nodes to aggregate multiple incoming stem transactions that arrive within a short period of time with one new transaction, then the adversary can reduce the number of newly generated transactions and hence the total cost of the attack.

\textbf{Attack is limited to transactions along stem paths.}
We note that this attack does not work on fluffed transactions. Let us consider the case that an adversarial node aggregates an incoming fluff transaction $T_A$ with a newly generated transaction $T_B$ and fluffs both $T_A + T_B$ and $T_B$. The fact that $T_A$ was sent to the adversarial node in the fluff phase means that some honest node(s) had $T_A$ as a fluff transaction. Since honest nodes follow the protocol and relay the incoming fluff transactions without aggregating them, $T_A$ will be broadcasted through the network. Hence, $T_A$ can still end up in the blockchain, even if miners prioritize $T_B$ over $T_A + T_B$.

\textbf{Attack mitigations.}
One way to mitigate this attack is to modify the wallet's source code to periodically resend previously broadcasted transactions that have not yet ended up in the blockchain. Increasing the number of retries for a transaction exponentially decreases the probability that the transaction does not appear in the blockchain. Nonetheless, even if we modify the wallet's source code, the attack can still significantly increase the latency for transactions. We also note that resending a transaction too frequently could cause the transaction to appear in multiple aggregations and therefore prevent other transactions from ending up in a block.

Another mitigation is to consider alternative routing protocols, such as those developed in the context of ad hoc networks, such as Castor~\cite{galuba2010castor}. In these protocols, each node keeps an estimate of reliability for each of its neighbors and makes routing decisions based on those estimates. We conjecture that there is value in adopting a similar idea to blockchain protocols concerned with source privacy. One design is for each node in the blockchain network to maintain an internal reliability score for each of its neighbors. The scores would be updated based on the feedback that nodes receive regarding the propagation of the transactions that they relayed to their neighbors. Using reliability scores, nodes can improve their relaying decisions. Scoring schemes have also been incorporated in other blockchain networks, such as Filecoin and Ethereum 2.0~\cite{vyzovitis2020gossipsub}.

We can also modify the protocol and disallow aggregation of transactions during the relay phase. For example, we could allow aggregation to occur only when transactions are being added to the blockchain. This approach mitigates the proposed denial of service attack but may compromise the content privacy of transactions as network nodes can observe transactions before they are aggregated with other transactions during the relay phase. Nonetheless, based on the information obtained from the log files of some Beam bootstrapping nodes\footnote{We have obtained information from the log files of bootstrapping nodes located in Europe-Frankfurt, USA-California, Hong Kong, and Singapore from the Beam developer community.} and also our main network node (deployed on University servers) from April 21, 2021, to April 23, 2021, we have observed that over $64\%$ of the incoming stem transactions to each of these nodes were not aggregated.


\textbf{Other attacks against MimbleWimble.}
Besides the denial of service attack described in this paper, we have designed and validated the modification of two well-known attacks~\cite{koshy2014analysis, fanti2018dandelion++}
against the Beam implementations of MimbleWimble. Here we briefly summarize our results. For more information about these attacks and our results please see the associated MSc thesis~\cite{tabatabaee2021attacking}.

\begin{enumerate}

    \item \textbf{Improved transaction source detection:} In this attack, the adversary uses information obtained from the content of incoming stem transactions for improved detection of the transaction source. We observed that the precision of the first node detection attack is $32\%$ for the single-kernel transactions while only $12\%$ for aggregated transactions. Therefore, performing the first node detection attack only on single-kernel transactions would lead to an improved transaction source detection.
    
    \item \textbf{Delaying transaction relay:} In this attack, to increase the latency of incoming transactions, the adversary adds excessive delays before forwarding stem transactions. We found that if $10\%$ of the network nodes are adversarial, by delaying transaction relay, the adversary can increase the expected transaction latency by over $31\%$.

\end{enumerate}

\textbf{Independent code review finds bugs.}
While studying the Beam source code, we found a validation bug in the function \emph{OnTransactionFluff} (Algorithm~\ref{alg:OnTransactionFluff}). This bug would allow an attacker to change the state of the stempool in an honest node by sending invalid transactions. 
%
%
We communicated with the Beam developer community about this issue and they modified the source code so that the function \emph{OnTransactionFluff} would validate transactions earlier\footnote{\url{https://github.com/BeamMW/beam/commit/ade19e1f8b1a702ad81d81092ba6a8f6561513ed}}.
To us, this experience further illustrates the value of independent code review for blockchain codebases.
\section{Conclusion}
\label{ch:conclusion}


MimbleWimble has been proposed as a cryptocurrency protocol that offers enhanced content privacy and prominent implementations of MimbleWimble, such as Beam, have adopted Dandelion++ for source privacy in transaction relay.

This paper contributes a transaction denial of service attack that uses MimbleWimble aggregation in combination with the Dandelion++ design. We evaluated this attack in a private test network of $100$ Beam nodes.
We found that if $10\%$ of the network nodes are adversarial, the attacker can prevent over $45\%$ of all transactions from ending up in the blockchain.
We presented several ideas for potential ways to mitigate this attack. We hope that our work will encourage more researchers to study MimbleWimble blockchains and their deployments.


%
\section*{Acknowledgment}
This work was supported by the Blockchain@UBC research cluster at the University of British Columbia, the Natural Sciences and Engineering Research Council of Canada (NSERC), Mitacs, and Aquanow. The authors would also like to thank the Beam developer community and Gleb Naumenko for their contributions to this project.

\bibliographystyle{IEEEtran}
\bibliography{IEEEabrv,biblio}

\begin{thebibliography}{10}
\providecommand{\url}[1]{#1}
\csname url@samestyle\endcsname
\providecommand{\newblock}{\relax}
\providecommand{\bibinfo}[2]{#2}
\providecommand{\BIBentrySTDinterwordspacing}{\spaceskip=0pt\relax}
\providecommand{\BIBentryALTinterwordstretchfactor}{4}
\providecommand{\BIBentryALTinterwordspacing}{\spaceskip=\fontdimen2\font plus
\BIBentryALTinterwordstretchfactor\fontdimen3\font minus
  \fontdimen4\font\relax}
\providecommand{\BIBforeignlanguage}[2]{{%
\expandafter\ifx\csname l@#1\endcsname\relax
\typeout{** WARNING: IEEEtran.bst: No hyphenation pattern has been}%
\typeout{** loaded for the language `#1'. Using the pattern for}%
\typeout{** the default language instead.}%
\else
\language=\csname l@#1\endcsname
\fi
#2}}
\providecommand{\BIBdecl}{\relax}
\BIBdecl

\bibitem{androulaki2013evaluating}
E.~Androulaki, G.~O. Karame, M.~Roeschlin, T.~Scherer, and S.~Capkun,
  ``Evaluating user privacy in bitcoin,'' in \emph{International Conference on
  Financial Cryptography and Data Security}.\hskip 1em plus 0.5em minus
  0.4em\relax Springer, 2013, pp. 34--51.

\bibitem{nakamoto2008bitcoin}
S.~Nakamoto, ``Bitcoin: A peer-to-peer electronic cash system,'' 2008.

\bibitem{meiklejohn2013fistful}
S.~Meiklejohn, M.~Pomarole, G.~Jordan, K.~Levchenko, D.~McCoy, G.~M. Voelker,
  and S.~Savage, ``A fistful of bitcoins: characterizing payments among men
  with no names,'' in \emph{Proceedings of the 2013 conference on Internet
  measurement conference}, 2013, pp. 127--140.

\bibitem{ober2013structure}
M.~Ober, S.~Katzenbeisser, and K.~Hamacher, ``Structure and anonymity of the
  bitcoin transaction graph,'' \emph{Future internet}, vol.~5, no.~2, pp.
  237--250, 2013.

\bibitem{ron2013quantitative}
D.~Ron and A.~Shamir, ``Quantitative analysis of the full bitcoin transaction
  graph,'' in \emph{International Conference on Financial Cryptography and Data
  Security}.\hskip 1em plus 0.5em minus 0.4em\relax Springer, 2013, pp. 6--24.

\bibitem{cryptoeprint2015monero}
S.~Noether, ``Ring signature confidential transactions for {M}onero,''
  Cryptology ePrint Archive, Report 2015/1098, 2015,
  https://eprint.iacr.org/2015/1098.

\bibitem{hopwood2016zcash}
D.~Hopwood, S.~Bowe, T.~Hornby, and N.~Wilcox, ``Zcash protocol
  specification,'' \emph{GitHub: San Francisco, CA, USA}, 2016.

\bibitem{poelstra2016mimblewimble}
A.~Poelstra, ``Mimblewimble,'' 2016.

\bibitem{rohrer2020counting}
E.~Rohrer and F.~Tschorsch, ``Counting down thunder: Timing attacks on privacy
  in payment channel networks,'' in \emph{Proceedings of the 2nd ACM Conference
  on Advances in Financial Technologies}, 2020, pp. 214--227.

\bibitem{bojja2017dandelion}
S.~Bojja~Venkatakrishnan, G.~Fanti, and P.~Viswanath, ``Dandelion: Redesigning
  the bitcoin network for anonymity,'' \emph{Proceedings of the ACM on
  Measurement and Analysis of Computing Systems}, vol.~1, no.~1, pp. 1--34,
  2017.

\bibitem{fanti2018dandelion++}
G.~Fanti, S.~B. Venkatakrishnan, S.~Bakshi, B.~Denby, S.~Bhargava, A.~Miller,
  and P.~Viswanath, ``Dandelion++ lightweight cryptocurrency networking with
  formal anonymity guarantees,'' \emph{Proceedings of the ACM on Measurement
  and Analysis of Computing Systems}, vol.~2, no.~2, pp. 1--35, 2018.

\bibitem{mitani2020anonymous}
T.~Mitani and A.~Otsuka, ``Anonymous probabilistic payment in payment hub,''
  2020.

\bibitem{confidentialtxs}
G.~Maxwell, ``Confidential transactions,'' 2016.

\bibitem{maxwell2013coinjoin}
------, ``Coinjoin: Bitcoin privacy for the real world,'' in \emph{Post on
  Bitcoin forum}, 2013.

\bibitem{grinnn}
``Grin developers, {G}rin,'' \url{https://grin-tech.org/}, accessed:
  2021-04-23.

\bibitem{beammm}
``Beam developers, {B}eam,'' \url{https://beam.mw/}, accessed: 2021-04-23.

\bibitem{heilman2015eclipse}
E.~Heilman, A.~Kendler, A.~Zohar, and S.~Goldberg, ``Eclipse attacks on
  bitcoin’s peer-to-peer network,'' in \emph{24th $\{$USENIX$\}$ Security
  Symposium ($\{$USENIX$\}$ Security 15)}, 2015, pp. 129--144.

\bibitem{biryukov2014deanonymisation}
A.~Biryukov, D.~Khovratovich, and I.~Pustogarov, ``Deanonymisation of clients
  in bitcoin p2p network,'' in \emph{Proceedings of the 2014 ACM SIGSAC
  Conference on Computer and Communications Security}, 2014, pp. 15--29.

\bibitem{fanti2017deanonymization}
G.~Fanti and P.~Viswanath, ``Deanonymization in the bitcoin p2p network,'' in
  \emph{Proceedings of the 31st International Conference on Neural Information
  Processing Systems}, 2017, pp. 1364--1373.

\bibitem{decker2014bitcoin}
C.~Decker and R.~Wattenhofer, ``Bitcoin transaction malleability and mtgox,''
  in \emph{European Symposium on Research in Computer Security}.\hskip 1em plus
  0.5em minus 0.4em\relax Springer, 2014, pp. 313--326.

\bibitem{andrychowicz2015malleability}
M.~Andrychowicz, S.~Dziembowski, D.~Malinowski, and {\L}.~Mazurek, ``On the
  malleability of bitcoin transactions,'' in \emph{International Conference on
  Financial Cryptography and Data Security}.\hskip 1em plus 0.5em minus
  0.4em\relax Springer, 2015, pp. 1--18.

\bibitem{bunz2018bulletproofs}
B.~B{\"u}nz, J.~Bootle, D.~Boneh, A.~Poelstra, P.~Wuille, and G.~Maxwell,
  ``Bulletproofs: Short proofs for confidential transactions and more,'' in
  \emph{2018 IEEE Symposium on Security and Privacy (SP)}.\hskip 1em plus 0.5em
  minus 0.4em\relax IEEE, 2018, pp. 315--334.

\bibitem{fuchsbauer2019aggregate}
G.~Fuchsbauer, M.~Orr{\`u}, and Y.~Seurin, ``Aggregate cash systems: A
  cryptographic investigation of mimblewimble,'' in \emph{Annual International
  Conference on the Theory and Applications of Cryptographic Techniques}.\hskip
  1em plus 0.5em minus 0.4em\relax Springer, 2019, pp. 657--689.

\bibitem{betarte2020towards}
G.~Betarte, M.~Cristi{\'a}, C.~Luna, A.~Silveira, and D.~Zanarini, ``Towards a
  formally verified implementation of the mimblewimble cryptocurrency
  protocol,'' in \emph{International Conference on Applied Cryptography and
  Network Security}.\hskip 1em plus 0.5em minus 0.4em\relax Springer, 2020, pp.
  3--23.

\bibitem{silveira2021formal}
A.~Silveira, G.~Betarte, M.~Cristi{\'a}, and C.~Luna, ``A formal analysis of
  the mimblewimble cryptocurrency protocol,'' \emph{arXiv preprint
  arXiv:2104.00822}, 2021.

\bibitem{galuba2010castor}
W.~Galuba, P.~Papadimitratos, M.~Poturalski, K.~Aberer, Z.~Despotovic, and
  W.~Kellerer, ``Castor: Scalable secure routing for ad hoc networks,'' in
  \emph{2010 Proceedings IEEE INFOCOM}.\hskip 1em plus 0.5em minus 0.4em\relax
  IEEE, 2010, pp. 1--9.

\bibitem{vyzovitis2020gossipsub}
D.~Vyzovitis, Y.~Napora, D.~McCormick, D.~Dias, and Y.~Psaras, ``Gossipsub:
  Attack-resilient message propagation in the filecoin and eth2. 0 networks,''
  \emph{arXiv preprint arXiv:2007.02754}, 2020.

\bibitem{koshy2014analysis}
P.~Koshy, D.~Koshy, and P.~McDaniel, ``An analysis of anonymity in bitcoin
  using p2p network traffic,'' in \emph{International Conference on Financial
  Cryptography and Data Security}.\hskip 1em plus 0.5em minus 0.4em\relax
  Springer, 2014, pp. 469--485.

\bibitem{tabatabaee2021attacking}
S.~A. Tabatabaee, ``Attacking transaction relay in mimblewimble blockchains,''
  Master's thesis, University of British Columbia, 2021.

\end{thebibliography}
\balance

\end{document}